\documentclass[conference]{IEEEtran}
\IEEEoverridecommandlockouts
\usepackage{cite}
\usepackage{amsmath,amssymb,amsfonts}
\usepackage{algorithmic}
\usepackage{graphicx}
\usepackage{subcaption}
\usepackage{xurl}
\usepackage{textcomp}
\usepackage{xcolor}
\usepackage{enumitem}
\usepackage{eso-pic}

\newcommand{\PlaceIEEEPermNotice}{
  \AddToShipoutPictureFG*{
    \AtTextLowerLeft{
      \raisebox{-0.55in}[0pt][0pt]{
        \makebox[\textwidth][c]{
          \begin{minipage}{0.97\textwidth}
          \centering
          \footnotesize
          \textcopyright~2026 IEEE. Personal use of this material is permitted. Permission from IEEE must be obtained for all other uses, in any current or future media, including reprinting/republishing this material for advertising or promotional purposes, creating new collective works, for resale or redistribution to servers or lists, or reuse of any copyrighted component of this work in other works.
          \end{minipage}
        }
      }
    }
  }
}

\def\BibTeX{{\rm B\kern-.05em{\sc i\kern-.025em b}\kern-.08em
    T\kern-.1667em\lower.7ex\hbox{E}\kern-.125emX}}
\begin{document}

\title{Secure Pseudonymetry with DSSS Watermarking\\for LEO Satellites
\thanks{This work was supported in part by NSF award 2348589.}
}

\author{Nisanur~Camuzcu and Alireza~Vahid
        \thanks{Nisanur Camuzcu and Alireza Vahid are with the Department of Electrical and Microelectronic Engineering, Rochester Institute of Technology, Rochester, NY 14623 USA (e-mails: nc1138@rit.edu; arveme@rit.edu).}
        }

\maketitle
\PlaceIEEEPermNotice

\begin{abstract}
Low-Earth-orbit (LEO) constellations offer dense signals of opportunity (SOP) for positioning, navigation, and timing (PNT) without dedicated navigation infrastructure. However, using these signals passively, by only listening to the downlink, is difficult when beacon/pilot components are weak, signals from multiple satellites overlap, or the receiver cannot reliably identify which satellite generated a given observable. This paper proposes a low-power direct-sequence spread-spectrum (DSSS) watermarking framework that embeds a recoverable pseudonymous satellite identifier and keyed authentication field into a LEO downlink signal. The watermark carries a public satellite pseudonym for identification and a keyed HMAC-SHA256-based authentication field for authorized verification, enabling passive detection and verification without full downlink demodulation. We evaluate the framework in a multi-satellite simulation using Starlink two-line-element (TLE)-derived geometry, Doppler, and link-budget scaling with Sionna RT-based local propagation. The results demonstrate candidate-satellite detection, public-ID recovery, and keyed verification under co-observed satellite interference, with limited perturbation to the simulated beacon/pilot and primary downlink components. The framework targets future transmitter-enabled LEO systems while remaining transparent to legacy receivers that do not process the low-power watermark.
\end{abstract}


\section{Introduction}

Reliable positioning, navigation, and timing (PNT) is critical in GNSS-denied or GNSS-degraded environments, including urban canyons, indoor/obstructed areas, disaster-response operations, and contested military scenarios. Low-Earth-orbit (LEO) broadband constellations provide a dense set of fast-moving transmitters with favorable geometry, strong Doppler diversity, and global coverage, making them attractive signals of opportunity for passive PNT~\cite{doppler_9652890}. In a passive setting, the receiver only listens to already transmitted satellite signals and does not require a dedicated navigation payload, a two-way link, or new terrestrial infrastructure.

Processing the full LEO downlink waveform may provide richer navigation observables by exploiting broader waveform and frame structure. However, this approach can require wideband sampling, detailed waveform knowledge, and frame-structure access that may not be available to a low-complexity passive PNT receiver~\cite{waveform_req_10107477}. In field-deployable or military-oriented settings, a passive receiver may instead rely on limited beacon/pilot guidance and satellite-specific identification or authentication observables while remaining receive-only.

Existing opportunistic LEO-PNT studies have therefore explored satellite-specific beacon, pilot, and reference components that can be observed without full payload decoding~\cite{beacon_jardak_s23063234, humphreys_2026}. These components are attractive because they can provide Doppler- and timing-related observables while avoiding full demodulation of the primary downlink waveform. For Starlink, narrowband beacon components near the downlink channel center have been reported as persistently observable during satellite passes~\cite{beacon_10149748}, and prior work has used such beacon-derived Doppler-rate measurements for passive LEO-aided PNT~\cite{vtc_2026}. However, reliable identification and attribution of these observables remains a key challenge, especially in future scenarios: the receiver must know which satellite generated the observable used in the navigation solution.

\begin{figure} [htbp]
    \centering \vspace{-0.2 cm}
    \includegraphics[width=0.95\linewidth]{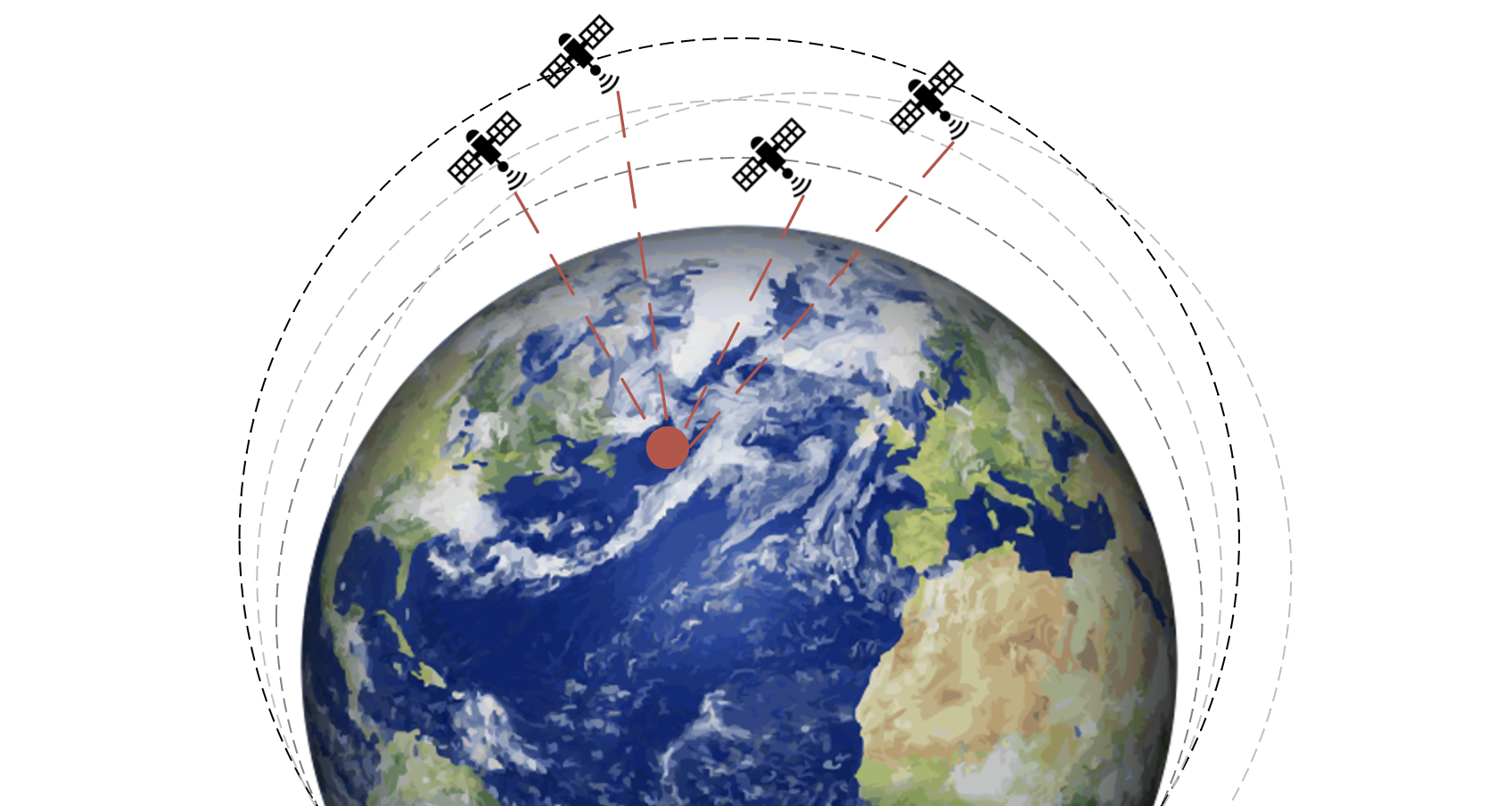}
    \caption{Dense multi-satellite future setting. \textit{The receiver only listens to satellite downlink components and must associate recovered observables with the correct satellite.}}
    \label{fig:future_setting}
\end{figure}

As illustrated in Fig.~\ref{fig:future_setting}, in emerging dense LEO deployments, multiple satellites are visible from the same receiver, and their reference components may appear in nearby time-frequency regions. Hence, a passive receiver that is not part of the network control protocol may observe a signal feature but may not reliably associate it with a specific satellite, beam, or transmitter. This ambiguity is problematic because the receiver must know which satellite generated the observable used in the navigation solution. Recent Starlink measurements further show that pilot-tone observability has degraded after 2023, making earlier pilot-tone-based approaches less reliable for low-gain passive receivers~\cite{kozhaya2025unveiling,humphreys_2026}. These limitations motivate a forward-compatible signal-layer enhancement that preserves passive reception while making satellite-specific observables easier to detect, identify, and verify.

A suitable enhancement must be low-power, recoverable through correlation, and independent of full payload demodulation. Direct-sequence spread-spectrum (DSSS) satisfies these requirements through spreading gain, which is central to GNSS acquisition~\cite{dsss_gps_6243707}, and has also been investigated in DSSS-overlaid OFDM receivers for joint communication and positioning~\cite{dsss_ofdm_11184853}. Recent work has further studied DSSS acquisition under high-dynamics LEO satellite communication~\cite{mcdsss_leo_11418619,dsss_leo_11122601}.

Motivated by these properties, we propose a DSSS pseudonymetry framework for passive LEO-aided PNT, in which a low-power watermark is superimposed on the existing downlink rather than transmitted as a separately allocated DSSS channel. Here, pseudonymetry refers to embedding a recoverable, replaceable satellite identifier into the transmitted signal for passive attribution. The watermark remains below the primary waveform and is recovered through correlation, enabling satellite identification and verification without full downlink demodulation, unlike a conventional header available only after successful frame acquisition and demodulation.

The watermark is a transmitter-side enhancement for future LEO systems and is not part of current Starlink transmissions, requiring operator participation for passive PNT/authenticated attribution while remaining transparent to legacy receivers. It is structured around three functions: correlation-based synchronization, public pseudonym identification, and keyed authentication using a hash-based message authentication code (HMAC). The public identification field enables satellite attribution, while the keyed authentication field allows authorized passive receivers to verify that the detected pseudonym was generated by a transmitter possessing the authorized key.

The design builds on prior pseudonymetry work that watermarks transmitted OFDM signals for passive emitter identification~\cite{pseudonymetry_10615923}, but targets LEO-aided PNT under high dynamics and multi-satellite coexistence. We therefore use DSSS pseudonymetry to denote its satellite-oriented physical-layer realization rather than a direct implementation of the original protocol. The receiver is motion-aware: Doppler and Doppler drift are derived from two-line-element (TLE)-based satellite-to-receiver range rate and range acceleration. A beacon-assisted front end provides coarse timing and frequency guidance, after which the DSSS watermark is used for synchronization, public-ID detection, and keyed verification.

We evaluate the proposed framework in a conservative multi-satellite simulation setting. Real Starlink TLEs are used to obtain slant range, path loss, delay, Doppler, and Doppler drift~\cite{celestrak_starlink_tle}, while a \textit{Sionna RT} campus scene provides local multipath and blockage through a near-receiver ray-tracing proxy~\cite{sionna}. The received signal includes the candidate satellite, co-observed satellites, a modeled primary downlink component, the low-power DSSS watermark, and noise. The watermark is not placed in a protected quiet subband of the primary downlink; instead, passive primary-waveform suppression is applied before DSSS acquisition. The simulation therefore tests whether a weak watermark can still be recovered and authenticated under multi-satellite coexistence.

The main contributions of this paper are as follows:
\begin{itemize} [leftmargin=*, itemsep=0.1em, topsep=0.1em]
    \vspace{-0.05 cm}\item We propose a DSSS pseudonymetry architecture for passive LEO-aided PNT, where future LEO systems embed a low-power transmitter-side spread-spectrum watermark into the primary downlink component.
    
    \item We design a watermark frame that separates public satellite identification from authorized authentication. A public DSSS field enables pseudonym detection, while an HMAC-SHA256-based DSSS field enables keyed verification by authorized passive receivers.
    
    \item We develop a motion-aware passive receiver that combines beacon-assisted timing/frequency guidance, DSSS acquisition/tracking, keyed verification, and primary-waveform suppression.
    
    \item We evaluate the framework in a multi-satellite simulation using real Starlink TLE-derived geometry and \textit{Sionna RT}-based local propagation. The results demonstrate low-power watermark recovery and successful authentication for a dish-assisted passive receiver.
\end{itemize}

\section{System Model}

We consider a passive receiver observing a limited baseband portion of a representative LEO downlink. Each satellite signal is modeled as the superposition of three components: (i) a primary downlink component occupying the observed band, (ii) a beacon/pilot component used for coarse acquisition and frequency guidance, and (iii) a low-power DSSS watermark used for pseudonymous transmitter identification and authentication. The transmitted complex baseband signal from satellite $m$ is written as
\begin{equation}
    s_m(t) =
    s_{\mathrm{p},m}(t)
    + s_{\mathrm{b},m}(t)
    + s_{\mathrm{wm},m}(t),
\end{equation}
where $s_{\mathrm{p},m}(t)$ denotes the modeled primary downlink component in the observed band, $s_{\mathrm{b},m}(t)$ denotes the beacon/pilot component, and $s_{\mathrm{wm},m}(t)$ denotes the proposed watermark.

\begin{figure} [htbp] 
    \centering
    \includegraphics[width=1\linewidth]{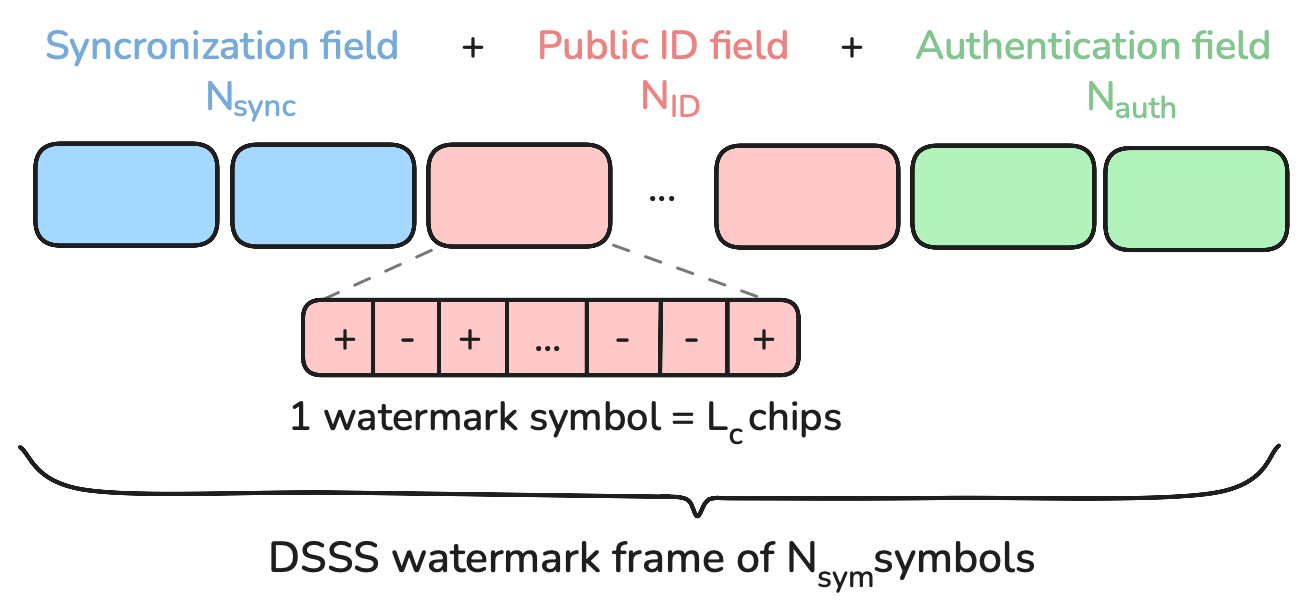}
    \caption{DSSS watermark frame structure. \textit{The frame contains synchronization, public-ID, and authentication fields, each composed of $N_{\rm sync}$, $N_{\rm ID}$, and $N_{\rm auth}$ watermark symbols respectively. Each watermark symbol is spread into a $L_c$-chip DSSS sequence before transmission.}}
    \label{fig:dsss}
\end{figure}

The watermark is modeled as a low-power DSSS component,
\begin{equation}
    s_{\mathrm{wm},m}(t)
    =
    \sqrt{P_{\mathrm{wm},m}}\,
    x_{k_m}(t)\,
    e^{j2\pi \Delta f_{\mathrm{wm}}t},
\end{equation}
where $P_{\mathrm{wm},m}$ is the watermark power, $k_m$ is the public pseudonym/code-index assigned to satellite $m$, $x_{k_m}(t)$ is formed by spreading each watermark symbol with the $L_c$-chip DSSS code indexed by $k_m$, and $\Delta f_{\mathrm{wm}}$ denotes the watermark frequency offset relative to the primary downlink reference. The receiver is assumed to have access to the valid pseudonym-to-satellite mapping or schedule for the authorized candidate set. The DSSS watermark is structured around three functions as shown in Fig.~\ref{fig:dsss}: correlation-based synchronization, public pseudonym identification, and keyed authentication using HMAC. Thus, the watermark provides a low-rate satellite-specific observable without requiring the receiver to demodulate the primary downlink component.

The receiver observes a multi-satellite received signal containing the candidate satellite and co-observed satellites:
\begin{equation}
    \begin{aligned}
    r&(t)
    =
    \underbrace{\sum_{\ell}
    h_{c,\ell}(t)\,
    s_c(t-\tau_{c,\ell}(t))\,
    e^{j2\pi f_{D,c}(t)t}}_{\text{candidate satellite}} \\
    &+
    \underbrace{\sum_{\substack{m=1 \\ m\neq c}}^{M}
    \sum_{\ell}
    h_{m,\ell}(t)\,
    s_m(t-\tau_{m,\ell}(t))\,
    e^{j2\pi f_{D,m}(t)t}}_{\text{co-observed satellites}}
    + \underbrace{w(t)}_{\text{noise}},
    \end{aligned}
\end{equation}
where the first term is the candidate satellite contribution, the second term represents co-observed satellites, and $w(t)$ is receiver noise. Here, $c$ denotes the candidate satellite, $M$ is the number of visible satellites, and $h_{m,\ell}(t)$ and $\tau_{m,\ell}(t)$ denote the complex propagation gain and delay of path $\ell$. The Doppler term $f_{D,m}(t)$ is satellite dependent. Since each $s_m(t)$ includes the primary downlink component, beacon/pilot component, and DSSS watermark, the candidate watermark is observed in the presence of co-satellite signals, primary downlink and beacon/pilot components, multipath, Doppler, and noise.

The receiver processing is designed for high-dynamics LEO observations, where Doppler and Doppler-induced signal distortion must be compensated before reliable correlation or tracking~\cite{doppler_9652890, doppler_leo_10017734}. For the candidate satellite $c$, the receiver first applies motion compensation using TLE-based Doppler and Doppler-rate terms:
\begin{equation}
    r_c(t)
    =
    r(t)
    \exp\!\left[
    -j2\pi
    \left(
    \hat f_{D,c} t
    + \frac{1}{2}\hat{\dot f}_{D,c}t^2
    \right)
    \right],
\end{equation}
where $\hat f_{D,c}$ and $\hat{\dot f}_{D,c}$ are obtained from the candidate satellite range rate and range acceleration. This step removes the dominant LEO motion trend and reduces the residual delay/Doppler search space. After this compensation, a beacon/pilot-aided estimator provides a residual CFO estimate $\hat f_0$, and the DSSS residual-frequency hypothesis $f$ is searched only over a small neighborhood around $\hat f_0$.

After coarse motion compensation and beacon-aided frequency centering, the watermark component is translated to baseband and filtered:
\begin{equation}
    z(t)
    =
    \mathrm{LPF}
    \left\{
    r_c(t)e^{-j2\pi \Delta f_{\mathrm{wm}}t}
    \right\}.
\end{equation}
Chip-rate samples are obtained by integrate-and-dump processing over chip intervals:
\begin{equation}
    z[n]
    =
    \int_{nT_c}^{(n+1)T_c} z(t)\,dt ,
\end{equation}
where $T_c = 1/R_c$ is the chip duration and $R_c$ is the chip rate.

DSSS acquisition is performed by correlating the chip-rate sequence with candidate spreading codes over delay and residual-frequency hypotheses. Its search cost scales with the number of code-index, delay, and frequency hypotheses, while the synchronization/public-ID observation interval is $T_{\rm acq}=(N_{\rm sync}+N_{\rm ID})L_cT_c$ and full keyed verification uses $T_{\rm pkt}=N_{\rm sym}L_cT_c$, where $N_{\rm sym}=N_{\rm sync}+N_{\rm ID}+N_{\rm auth}$. For candidate code-index hypothesis $k$, chip delay $d$, and residual frequency $f$, the acquisition metric is
\begin{equation}
    \Lambda(k,d,f)
    =
    \sum_{b}
    \left|
    \sum_{n \in \mathcal{B}_b}
    z[n+d]\,
    c_k^*[n]\,
    e^{-j2\pi f nT_c}
    \right|,
\end{equation}
where $\mathcal{B}_b$ denotes a short group of consecutive chips over which the residual phase is assumed approximately constant. The receiver correlates coherently within each block and sums block magnitudes noncoherently across blocks, improving robustness to residual Doppler drift. This follows standard DSSS acquisition practice, where correlation-based searches and noncoherent accumulation are used for DSSS acquisition~\cite{dsss_acquisition_2793}. The detected public pseudonymous index is
\begin{equation}
    \hat{k}
    =
    \arg\max_{k,d,f}
    \Lambda(k,d,f).
\end{equation}

The authentication field is generated by applying HMAC-SHA256 to a frame message containing the public pseudonym and frame counter, using a secret key provisioned to authorized receivers. Checking the expected frame counter can reject previously recorded authentication fields; without such validation, replay of a previously valid frame remains possible. Key distribution, counter management, and pseudonym scheduling are treated as operator-provisioned functions outside the physical-layer receiver design. HMAC-SHA256 provides a standard keyed message-authentication construction, unlike a plain public hash that can be recomputed by any observer~\cite{hmac_11291474, rfc2104_hmac}. Thus, an attacker knowing only the public codebook can identify or imitate a public pseudonym, but cannot generate a valid authentication tag for that pseudonym-counter combination without the secret key. The tag bits are carried by DSSS-spread authentication symbols and verified by comparing the recovered bits with the expected HMAC output.

\section{Simulation Setup}

\begin{figure} [t]
    \centering
    \includegraphics[width=1\linewidth]{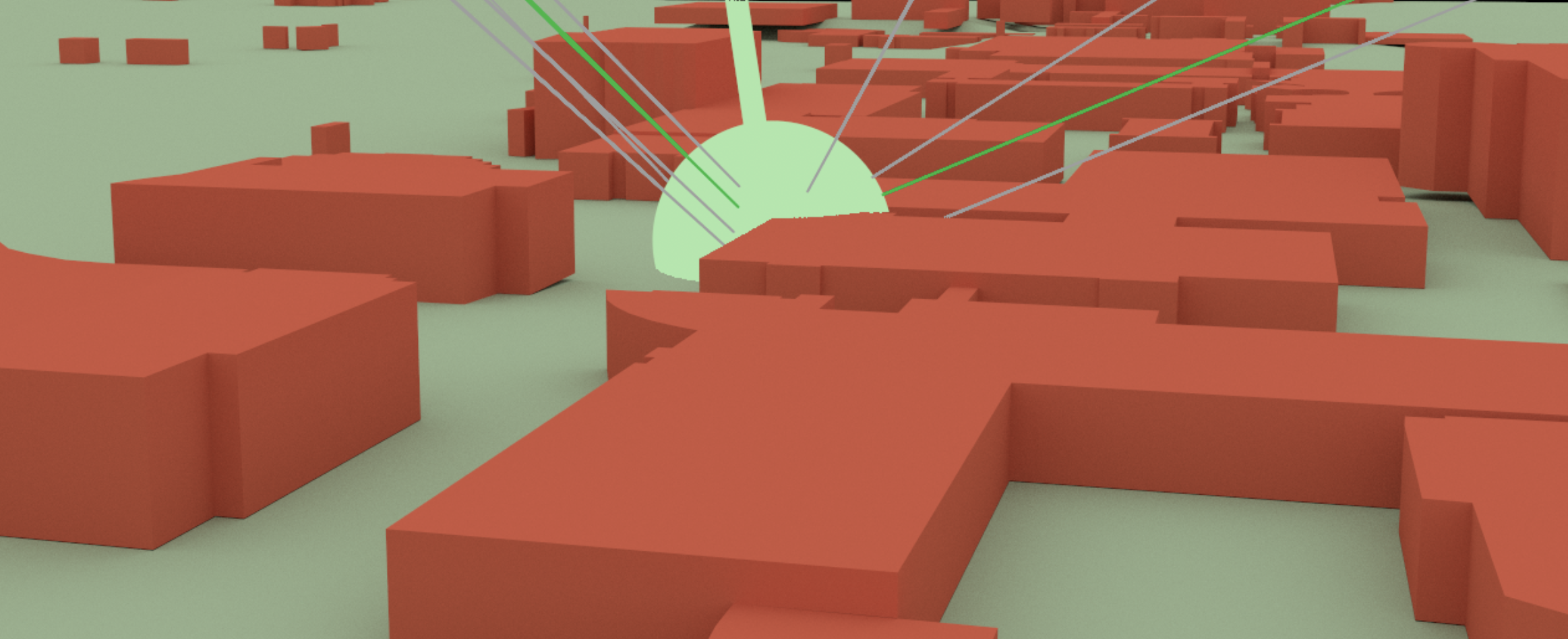}
    \caption{Sionna RT scene of the RIT campus used to model local propagation around the passive receiver. \textit{The light green sphere represents the receiver, while green and gray rays indicate LoS and NLoS propagation paths.}}
    \label{fig:rays}
\end{figure}

The simulation is implemented in Python 3.14 and uses a post-2023-inspired LEO downlink model observed over a limited baseband bandwidth. Starlink TLEs processed with Skyfield provide satellite geometry, slant range, free-space path loss (FSPL), delay, Doppler, and Doppler drift. Local receiver-side propagation is modeled using a Sionna RT scene of the RIT campus, as shown in Fig.~\ref{fig:rays}~\cite{sionna}: because the satellite lies outside the finite scene, a proxy transmitter is placed along its TLE-derived azimuth/elevation direction, and the resulting LoS/NLoS multipath is combined with the satellite-scale TLE quantities.

We evaluate a multi-satellite same-band observation scenario in which a single receiver snapshot contains the superposition of co-visible satellite signals. The components share the observation interval but are not forced to be chip-synchronous; each satellite contribution follows its own modeled propagation delay, Doppler shift, and channel response. In the representative run, the closest interferer is separated from the candidate by about $77~\mu$s, or roughly $31$ DSSS chips, so the test represents asynchronous same-band overlap rather than a worst-case chip-aligned collision. TLE assistance provides the coarse candidate delay, Doppler, and Doppler-rate prediction, while the beacon/pilot component is used to refine the residual frequency offset before DSSS acquisition. 
The keyed authentication mechanism itself does not rely on TLEs; however, without TLE or ephemeris assistance, the receiver would require a wider blind search over delay and Doppler hypotheses, increasing acquisition complexity and potentially reducing detection reliability.

Received powers are set through a Ku-band link budget with TLE-derived FSPL, receiver gain, noise figure, bandwidth, and fixed relative signal offsets. In the simulation figures, ``host'' refers to the modeled primary downlink component on which the DSSS watermark is superimposed. The beacon/pilot and DSSS watermark are placed $25$ dB and $35$ dB below the primary downlink component, respectively. Thus, the watermark is not directly distinguishable in the composite received spectrum before despreading; it is recovered through DSSS processing gain and correlation-based acquisition. Table \ref{tab:sim_params} lists the principal receiver, propagation, and watermark design parameters used in the simulation. Geometry-dependent quantities are derived from the selected TLE scene, while chip rate, code length, watermark offset, and authentication-field structure are design parameters of the proposed watermark.

\begin{table}[t]
\centering
\caption{Main simulation and watermark parameters.}
\label{tab:sim_params}
\begin{tabular}{l c l c}
\hline
\textbf{Parameter} & \textbf{Value} & \textbf{Parameter} & \textbf{Value} \\
\hline
Carrier freq. & $11.325$ GHz & Observed bandwidth & $1$ MHz \\
Chip rate $R_c$ & $400$ kchips/s & Samples/chip & $6$ \\
Code length $L_c$ & $511$ chips & No. satellites & $8$ \\
$N_{\rm sync}$ & $48$ & $N_{\rm ID}$ & $12$ \\
$N_{\rm auth}$ & $68$ & $N_{\rm sym}$ & $128$ \\
$T_{\rm pkt}$ & $163.5$ ms & $\Delta f_{\rm wm}$ & $65.9$ kHz \\
Rx gain & $35$ dBi & Noise figure & $3$ dB \\
Beacon offset & $-25$ dB & Watermark offset & $-35$ dB \\
\hline
\end{tabular}
\vspace{-0.2 cm}
\end{table}

A chip is one element of the DSSS spreading sequence; in the simulation, each watermark symbol is spread over $L_c=511$ chips and $\Delta f_{\rm wm}\approx65.9$ kHz to avoid direct overlap with the beacon/pilot comb. The authentication field contains not only keyed authentication bits but also known preamble and pilot symbols for receiver tracking and equalization. The full watermark frame duration is $T_{\rm pkt}=N_{\rm sym}L_c/R_c$ for the proof-of-concept parameter set. Shorter frames can be obtained through alternative frame designs, higher chip rates, or more compact ID/auth fields. The reported frame settings balance packet duration and compatibility with existing observables, while TLE-derived motion compensation supports DSSS acquisition over this interval.

\section{Results}
In this section, we evaluate whether the proposed DSSS watermark can be detected, identified, and authenticated in a representative dish-assisted future multi-satellite setting. The results focus on three aspects: (i) detection and keyed verification under different watermark power and coexistence load conditions, (ii) candidate-satellite attribution when co-observed watermark components are present, and (iii) the impact of the watermark on existing beacon/pilot and primary downlink observables. Each sweep point is estimated from 100 independent Monte Carlo trials with randomized receiver noise and waveform initialization, and 95\% binomial confidence intervals are shown to indicate finite-sample uncertainty.

\subsection{Watermark Detection and Keyed Verification}

\begin{figure} [t]
    \centering
    \includegraphics[width=\linewidth]{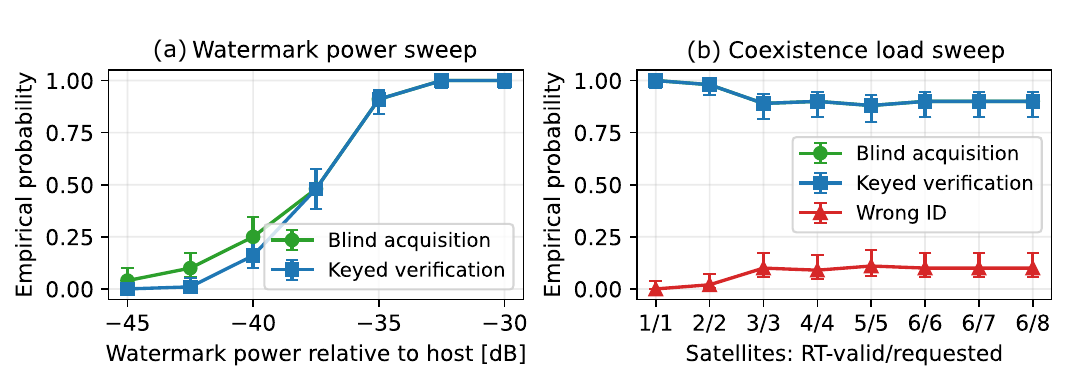}
    \caption{Watermark detection and keyed verification under power and coexistence sweeps. \textit{Plot (a) shows blind acquisition and keyed verification probability as the DSSS watermark power is varied relative to the primary downlink component, denoted as the host. Plot (b) shows candidate-satellite detection as the number of co-observed satellites increases in the multi-satellite setting. The results show a clear operating threshold in watermark power and sustained keyed verification performance under the tested coexistence load, with low wrong-ID probability.}}
    \vspace{-0.4 cm}
    \label{fig:coexistence}
\end{figure}

The first objective is to determine the operating region in which the watermark is strong enough to support reliable acquisition and authentication, while still remaining low-power relative to the primary downlink component. Therefore, Fig.~\ref{fig:coexistence} shows the sensitivity of the proposed receiver to DSSS watermark power and multi-satellite coexistence. In the watermark power sweep, blind acquisition and keyed verification remain unreliable when the watermark is below about $-37.5$ dB relative to the primary downlink component. Around the selected $-35$ dB operating region, both probabilities rise sharply toward near-unity success, indicating a clear threshold for reliable recovery. This behavior is expected because the watermark is recovered through correlation-based DSSS processing gain rather than direct spectral visibility.

The coexistence sweep in Fig.~\ref{fig:coexistence} evaluates candidate-satellite detection as additional co-observed satellites are included in the same TLE scene. The receiver maintains high keyed verification probability under the tested coexistence load, while the wrong-ID probability remains low. This result indicates that the public-ID and keyed verification stages can support candidate-satellite detection even when other satellite watermark components are present in the observed band.

\begin{figure} [b!]
    \centering
    \vspace{-0.3 cm} 
    \includegraphics[width=\linewidth]{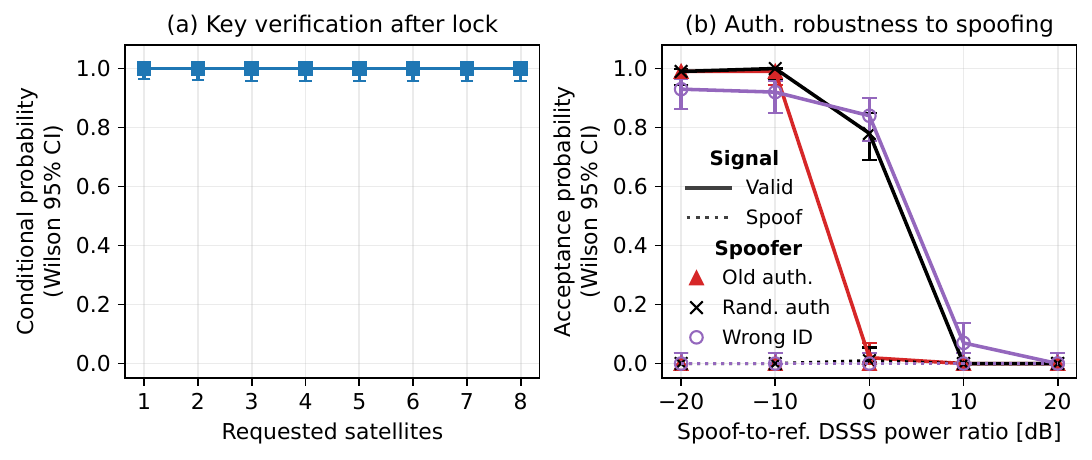}
    \vspace{-0.4 cm}
    \caption{Authentication stress tests. Plot (a) shows keyed verification after lock as the requested satellite count increases. Plot (b) compares valid-signal and spoof acceptance versus spoof-to-reference DSSS power for replayed/old-authentication, random-authentication, and wrong-ID attacks.}
    \label{fig:security}
\end{figure}

Fig.~\ref{fig:security} shows that the proposed authentication layer remains stable under increased multi-satellite demand while retaining strong selectivity against invalid authentication material. The near-unity verification rate after lock indicates that requesting additional satellites does not introduce a measurable keyed verification bottleneck in the tested scenario. The spoofing power sweep further separates valid signal acceptance from spoof false acceptance. Solid curves show that acceptance of the valid signal remains high when the spoofing signal is weak, but degrades once old-authentication, random-authentication, or wrong-ID spoofers approach or exceed the valid DSSS power. In contrast, the dotted curves show no invalid spoofing strategy produces spoof acceptance over the tested power range. Thus, invalid authentication material is rejected in the tested attacks, while sufficiently strong interference or spoofing can still prevent acquisition of the legitimate watermark. The keyed authentication layer therefore supports source verification but does not prevent denial of service through strong interference or spoofing, and replay protection depends on validation of the expected frame counter. 

\subsection{Candidate Attribution in a Multi-Satellite Setting}

\begin{figure} [b!]
    \centering
    \vspace{-0.2 cm} \includegraphics[width=\linewidth]{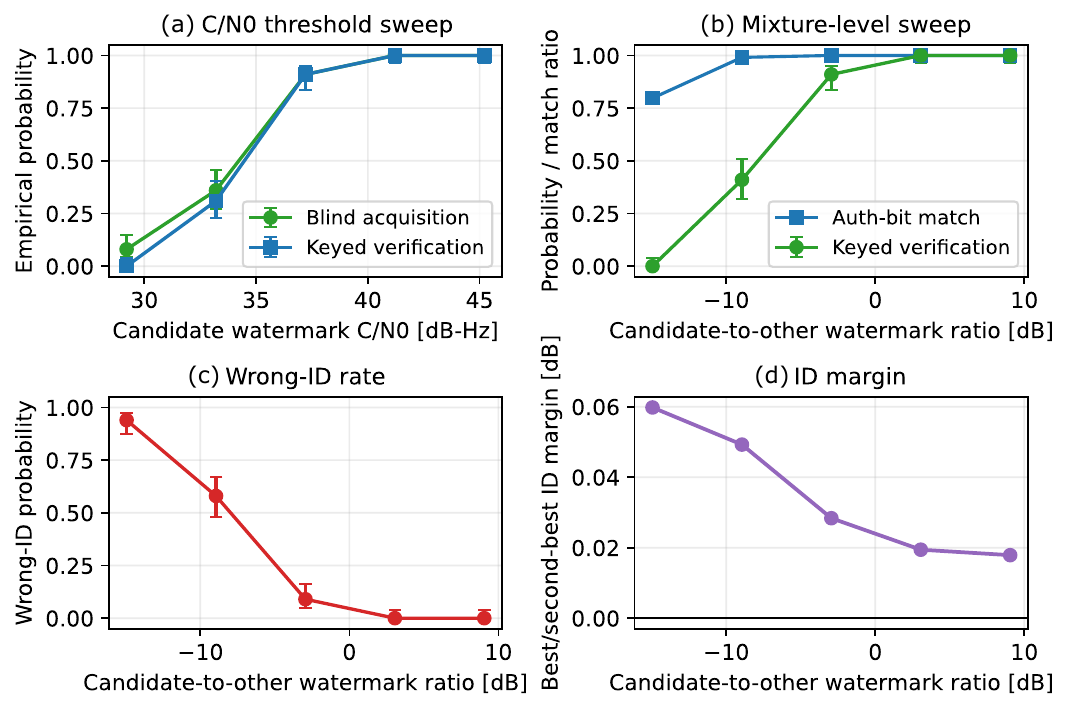}
    \caption{Candidate-satellite attribution in a multi-satellite setting. \textit{Plot (a) shows that increasing candidate watermark C/N0 improves blind acquisition and keyed verification. Plot (b) shows that increasing the candidate-to-other watermark ratio improves authentication-bit recovery and keyed verification. Plot (c) shows the corresponding reduction in wrong-ID probability, while Plot (d) reports the separation between the strongest and second-strongest public-ID hypotheses.}}
    \label{fig:detection}
\end{figure}

The next objective is to evaluate whether the receiver can attribute the recovered watermark to the correct candidate satellite when other satellites cause interference. Fig.~\ref{fig:detection} examines this attribution behavior under different candidate watermark carrier-to-noise-density ratio ($C/N_0$) and candidate-to-other watermark ratios. Here, $C/N_0$ quantifies the candidate watermark strength relative to receiver noise density before despreading, making it a useful weak-signal acquisition metric. As $C/N_0$ increases, both blind acquisition and keyed verification improve. This confirms that the receiver performance is primarily limited by the post-despreading reliability of the candidate watermark.

The mixture-level sweep evaluates the effect of the candidate-to-other watermark ratio. When the candidate watermark is weak relative to co-observed watermark components, the authentication-bit match and keyed verification probabilities are reduced, and wrong-ID events are more likely. As the candidate-to-other ratio increases, authentication-bit recovery improves and the wrong-ID probability decreases. Therefore, the receiver is not simply detecting the presence of a watermark; it is also able to associate the recovered watermark with the correct candidate satellite when sufficient separation exists between the candidate and co-observed components. The ID-margin plot reports the separation between the strongest and second-strongest public-ID hypotheses and provides an additional diagnostic of this attribution process.

\subsection{Impact on Existing Downlink Observables}

\begin{figure} [htbp]
    \centering
    \begin{subfigure}{\linewidth}
        \centering
        \includegraphics[width=\linewidth]{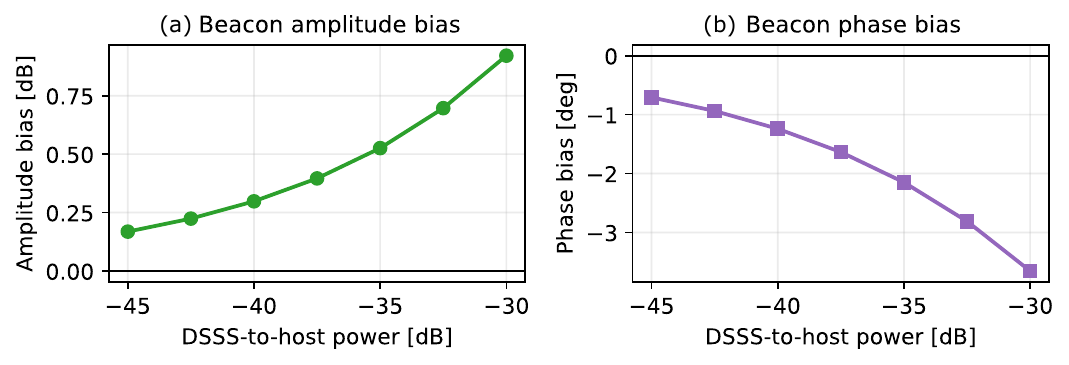}
    \end{subfigure}

    \begin{subfigure}{\linewidth}
        \centering
        \includegraphics[width=0.47\linewidth]{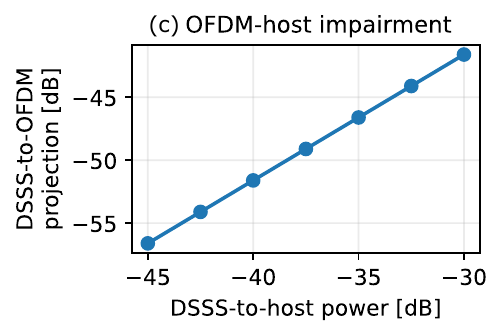}
    \end{subfigure}

    \caption{Impact of the DSSS watermark on existing downlink observables. \textit{Plots (a) and (b) show beacon amplitude and phase bias relative to the no-watermark reference as DSSS-to-primary power is varied. Stronger watermark power increases the bias magnitude, showing the tradeoff between watermark recoverability and compatibility with existing beacon/pilot observables. Plot (c) shows the projected distortion of the modeled primary waveform due to the watermark.}}
    \label{fig:compatibility}
\end{figure}

We also evaluate whether the low-power DSSS watermark remains compatible with existing downlink observables. For a receiver that does not despread the watermark, the DSSS layer is not recovered as a separate signal and instead appears as an additional low-power interference term. We therefore assess compatibility by measuring the watermark-induced changes in the beacon/pilot amplitude and phase estimates, as well as the projected distortion on the modeled primary downlink waveform, relative to the no-watermark case.
Fig.~\ref{fig:compatibility} shows that the watermark introduces a measurable but limited perturbation to the beacon/pilot observables, while its projection onto the modeled primary waveform remains far below the host. At the selected $-35$ dB operating point, the amplitude and phase biases remain moderate, while stronger watermark settings increase the perturbation. The beacon amplitude bias is $\sim0.5$ dB and the phase-bias magnitude is about $2^\circ$; the projected primary-waveform distortion is $-46.61$ dB, implying $0.00009$ dB equivalent SNR loss and $0.0013$ dB total power change.

\section{Conclusion}
This paper proposed a low-power DSSS watermarking framework for secure pseudonymetry in passive LEO-aided PNT. The watermark is superimposed below the primary downlink and recovered through DSSS correlation, carrying a public satellite pseudonym and an HMAC-SHA256-based authentication field for identification and keyed verification without full downlink demodulation. A motion-aware system was developed using TLE-assisted Doppler compensation, beacon/pilot-aided residual correction, DSSS acquisition, and authentication checking. Simulation results in a multi-satellite setting showed that the proposed watermark can support satellite detection, public-ID recovery, and keyed verification under co-observed satellite interference. The results also showed a compatibility tradeoff: increasing watermark power improves recoverability, while also increasing perturbation to existing beacon/pilot observables. Future work will optimize the watermark frame length and power allocation, evaluate smaller-aperture and bare-LNB receivers, study ephemeris uncertainty and validate the proposed design with hardware-calibrated LEO signal measurements.

\bibliographystyle{IEEEtran}
\bibliography{references}

\end{document}